\documentclass[twocolumn]{biophys-new}
\usepackage[utf8]{inputenc}
\usepackage{graphicx}
\usepackage[colorlinks,allcolors=cyan!70!black]{hyperref}
\usepackage{lipsum}

\title{Telomeres in Lamin-A Depleted Cells Exhibit Directed Motion and Dynamic Coherence}
\runningtitle{Biophysical Journal Template} 

\author[1,3,*]{Mario Hidalgo-Soria}
\author[2]{Wajdi Nicola}
\author[3]{Yifat Haddad}
\author[3]{Eli Barkai}
\author[3]{Stanislav Burov}
\author[4]{Yuval Garini}
\runningauthor{Mario Hidalgo-Soria et al} 

\affil[1]{Department of Biotechnology and Pharmaceutical Sciences, Western University of Health Sciences, Pomona, CA 91766, USA}
\affil[2]{Faculty of Biomedical Engineering \& , Russell Berrie Nanotechnology Institute, Technion, Israel Institute of Technology, Haifa 3200003, Israel}
\affil[3]{Department of Physics and Institute of Nanotechnology and Advanced Materials, Bar-Ilan University, Ramat-Gan 5290002, Israel}
\affil[4]{Faculty of Biomedical Engineering, Technion, Israel Institute of Technology, Haifa 3200003, Israel}

\corrauthor[*]{mariohidalgosoria@gmail.com}
\papertype{Article}

\begin{document}

\begin{frontmatter}

\begin{abstract}

Investigating the dynamics of chromatin loci and the factors that influence them provides valuable insights into the organization and functionality of the genome within the cell nucleus. We control the expression of Lamin-A, an important organizer of chromatin and nuclear structure. By simultaneously tracking hundreds of telomeres in Lamin-A knocked-out (KO) and wild-type (WT) nuclei, we find that telomere motion in Lamin-A depleted cells is both faster and more directed on micrometer scales, comparable to the size of chromosome territories. In contrast, telomere trajectories in WT cells exhibit pronounced anti-persistent behavior, consistent with caging by the surrounding chromatin environment. We further observe correlated motion between distinct telomeres in both WT and KO cells, with significantly stronger correlations in the KO case, indicating enhanced collective behavior. These correlations reflect cross-correlations among different loci rather than temporal correlations along individual trajectories. Together, these findings highlight the central role of Lamin-A in regulating both local confinement and collective telomere dynamics.
\end{abstract}

\begin{sigstatement}
  The study of telomere dynamics is determinant for identifying the main
  \  contributors to cellular stability, aging,  chromatin integrity and related diseases.
  We reveal the essential role of Lamin-A in telomere dynamics by simultaneously tracking their trajectories in live cells.
  Lamin-A depletion increases telomere spreading and directional movement at the scale of chromosome territories.
  Additionally, neighboring telomeres exhibit stronger correlations, linked to the  directed nature of their trajectories.
  These findings unravel Lamin-A’s role in maintaining nuclear organization and genome stability.
\end{sigstatement}
\end{frontmatter}

\section*{Introduction}

$ \,\ $ Chromatin dynamics play a central role in regulating key nuclear processes ~\cite{Zakian1995,Bohr1995}, as the spatial and temporal mobility of genomic loci directly influences gene expression, DNA repair, and overall genome stability. The movement of chromatin modulates the accessibility and interactions of regulatory entities with the transcriptional machinery, thereby affecting its efficiency. In parallel, constrained chromatin dynamics facilitate the localization of DNA repair factors thereby promoting efficient and timely repair. Importantly, alterations in chromatin dynamics have been linked to pathological conditions ~\cite{misteli2020}. Thus, investigating chromatin dynamics provides critical insight into how nuclear organization and physical constraints govern genome function and stability.

Given its importance, chromatin dynamics has been extensively studied. Its dynamics is regulated by nuclear proteins, among which Lamin-A plays a central role, but other proteins are important as well ~\cite{{Viva2019}}. Together with Lamin-B, Lamin-A forms the nuclear lamina, a meshwork lining the inner nuclear envelope that provides mechanical stability and contributes to gene regulation. In addition to this structural role, a substantial fraction of Lamin-A resides within the nuclear interior, where it interacts with chromatin. Dysregulation of Lamin-A expression is associated with a range of pathological conditions, collectively termed laminopathies ~\cite{gilchrist2004}, and its deficiency has been shown to increase chromatin mobility, likely impacting chromatin function ~\cite{Vered2008,Bronshtein2015,prokocimer2009,de2010increased,melcer2012histone,melcer2012,mahen2013}.

Chromatin dynamics have been investigated using multiple approaches, including fluorescence recovery after photobleaching ~\cite{melcer2012histone}, displacement correlation spectroscopy (DCS) \cite{Zidovska2013}, Hi-C ~\cite{samejima2025}, and single-particle tracking of nuclear structures such as telomeres and nucleosomes ~\cite{Bronshtein2015,mahen2013,BornGB2009,kepten2015uniform,iida2022}. Telomeres, DNA–protein complexes at chromosome ends, serve as well-defined and trackable chromatin loci, making them reliable reporters of global chromatin behavior. Labeling telomere-associated proteins (e.g., TRF1) provides bright, stable signals that enable high-precision tracking over extended periods. In addition, their abundance (92 in human cells, 80 in mice) and distribution throughout the nuclear volume allow probing chromatin dynamics across large spatial scales.

To extract the biological significance of chromatin dynamics, it is essential to identify appropriate biophysical models and its interpretation.  Telomere motion has often been reported to exhibit sub-diffusive behavior ~\cite{BornGB2009,kepten2015uniform}, commonly interpreted within the framework of the de Gennes reptation model~\cite{deGennes1971}. This interpretation attributes the observed dynamics to caging by the surrounding chromatin environment. However, sub-diffusion alone does not, in general, imply caging. Moreover, active processes within the nucleus can generate persistent or directed motion ~\cite{CHUANG2006}, leading to qualitatively distinct dynamical regimes. Throughout this study, the terms persistent and directed are used interchangeably. A central challenge is therefore to quantify persistence in chromatin motion and to identify experimentally controllable parameters that govern transitions between caged and persistently driven dynamic.

Several studies have shown that telomere and chromatin dynamics are strongly influenced by active, non equilibrium forces. For example, Weiss  \textit{et al.} ~\cite{Stadler2017} reported that disrupting micro-tubules reduces telomere diffusivity, demonstrating that sub-diffusive telomere motion in mammalian nuclei depends on cytoskeleton-generated active forces. Similarly, Makhĳa et al.~\cite{Makhija2016} observed on artificially deformed and polarize-constraint NIH 3T3 fibroblast cells that Lamin-A/C deficiency, reduces significantly the pair-correlation of telomere motion, while it significantly increases the nuclear area fluctuations. The pair-correlation of telomeres in these cells was also significantly reduced when treating it with cytochalasin-D that perturb the myosin activity. Together, these findings suggest that active cytoskeletal forces, in concert with Lamin-A/C mediated nuclear rigidity, regulate chromatin motion and dynamics.

Additional studies have identified active processes operating within the nucleus. Using Hi-C, Mirny et al. ~\cite{samejima2025} showed that cohesin-mediated loop extrusion generates forces that reposition genomic elements and modulate promoter–enhancer interactions. Meanwhile, using DCS, Zidovska et al. \cite{Zidovska2013} revealed micron-scale coherent chromatin motion  $\sim 4-5 \mu m$ persisting for several seconds, attributed to ATP-dependent activity.

However, DCS describes bulk chromatin motion and do not directly quantify the relative dynamics between specific genomic loci. Conversely, single particle tracking studies often focus on mean-squared displacement or time-averaged quantities, which overlook spatial correlations between loci \cite{Bronshtein2015,Shinkai2016,Lucas2014}. Thus, a quantitative description of dynamic correlations between individual chromatin loci remains lacking.

Here, we address this gap by measuring telomere dynamics in live cells. We analyze pairwise temporal and spatial correlations as a function of inter-loci distance, enabling quantification of spatial coherence and collective chromatin motion at the single-locus level. We further examined telomere directionality to identify preferred orientations and developed a bimodal model capturing two distinct dynamical regimes. Our results reveal a dominant role for Lamin-A in suppressing micrometer-scale directional motion, thereby preserving chromatin integrity and limiting large local displacements that may lead to chromosomal intermixing or aberrations. Consistently, Lamin-A deficiency leads to more directed and persistent telomere motion.

We analyze telomere dynamics in two types of mouse embryonic fibroblasts (MEFs): wild-type (WT) cells expressing Lamin-A and Lamin-A knockout (KO) cells. Telomere motion in KO cells is significantly enhanced compared to WT, consistent with previous studies ~\cite{Makhija2016}. In the absence of Lamin-A, telomeres exhibit persistent motion over substantial distances, reaching up to $\sim20$\% of the nuclear diameter. We further show that pairwise correlations between loci are linked to this persistent motion. These findings highlight the key role of Lamin-A in regulating telomere dynamics and suppressing large-scale chromatin motion within the nucleus.

\section*{Materials And Methods}
\subsection*{\label{sec:21}\textbf{Experimental setup}}

\begin{figure*}
\begin{center}
  \includegraphics[width=0.9\textwidth]{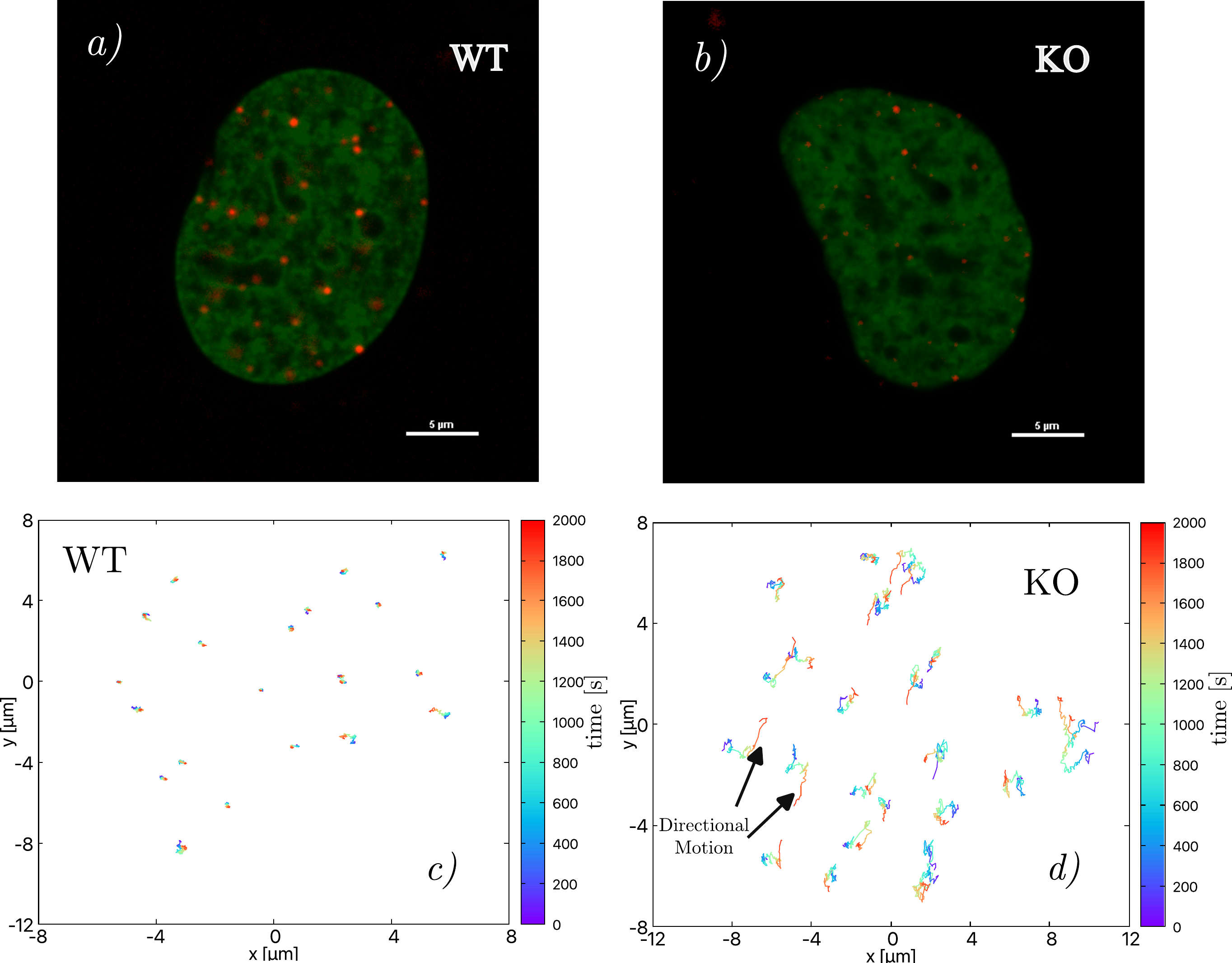}
\end{center}
  \vspace{-6mm}
  \caption{{\protect\footnotesize {  Microscopic images of the cell nucleus showing telomeres in red and H2B histones in green, scale bar $5 \mu m$. \textit{a)} WT cell. \textit{b)} KO cell. Trajectories of telomeres in the nucleus. \textit{c}) Telomeres path in  wild type (WT) cells. \textit{d}) Telomeres path in Lamin-A depleted cells (KO). Telomeres moving in cells depleted of Lamin-A cover larger distances compared with telomeres diffusing in WT cells, and directional motion often occurs. The color gradient represents time in seconds. See Fig.~S6 in SM for a detailed description of telomere tracks in WT cells.
  }}}
\vspace{0mm}
  \label{fig:micrograph}
\end{figure*}

Measurements were carried out in both WT and KO MEFs~\cite{sullivan1999}, and cell lines were kindly provided by Prof. Susana Gonzalo from Saint Louis University School of Medicine, St Louis, MO, USA.
The cells were maintained in Dulbecco’s high-glucose modified Eagle’s medium containing 10\% fetal bovine serum, 1\% penicillin and streptomycin antibiotics and 1\% L-glutamine, all materials are from Biological Industries, Beit HaEmek, Israel. For validating the measured dynamic properties, we tracked telomeres dynamics in two different experimental setups.

In the first setup 13 WT  and 7 KO cells  were analyzed, yielding 363 WT  and 128  KO telomere trajectories. We used transient labeling of the shelterin subunit TRF2 fused to green fluorescent protein (GFP), a kind gift from Prof. Sabine Mai (Manitoba Institute of Cell Biology, Winnipeg University, Canada). The imaging system includes an inverted Olympus IX-81 fluorescence microscope coupled to an FV-1000 confocal set-up (Olympus, Tokyo, Japan), and a UPLSAPO X60 objective lens (NA=1.35). Cells were placed in a $37^{\circ}$C incubator (Tokai, Shizuoka-ken, Japan) with a 5\% $CO_2$ level. The measurements were performed in three dimensions to correct for cell linear and rotational drift, although the actual dynamic analysis was performed only on the planar $xy$ motion, see details below. Normally 35 equally spaced planes were measured with a lag time of 18.5 seconds, a voxel size of 108 X 108 X 350 $nm^{3}$ and 50 or 100 time-points. The planar resolution of our measurements  is ~170 nm, and the spatial precision is better than 10 nm (localization error), based on both calculation and measurement of a fixed sample.

In the second setup, 6 WT and 5 KO cells were analyzed, with 234 WT and 248 KO tracks. We employed transient labeling of TRF1 with the red florescent protein DsRed (see Fig.~\ref{fig:micrograph}~\textit{a-b)} and supplementary videos S1 and S2), a kind gift from Vered Raz (Leiden University Medical Center, The Netherlands). Imaging was carried out on a Nikon Ti2-E spinning-disk confocal microscope, which enables fast acquisition with minimal phototoxicity, using a CFI SR HP PLAN APOCHROMAT LAMBDA S X100 objective lens (NA=1.35). Cells were maintained at $37^{\circ}$C incubator with 5\% $CO_2$ level throughout imaging. Three-dimensional stacks were acquired to correct for both translational and rotational drift. For each cell, 41 equally spaced Z-planes were recorded with a lag time of 18.5 seconds and a 100 ms exposure, providing a robust signal-to-noise ratio. Telomere positions were extracted using Imaris for single particle tracking, followed by further analysis using our own software tools.

\subsection*{\label{sec:221}\textbf{  Correction of linear and rotational drift.}}

As mentioned above we employed confocal microscopy to obtain 3D data sets of telomere's trajectories. As the cells are alive, they tend to move during the data acquisition. In order to describe the motion of telomeres with respect to the nucleus, the nucleus motion has to be subtracted from the telomere's trajectories. Indeed, the observation of telomere dynamics before and after corrections reveals that it is necessary to rectify the translational drift, and also but less importantly the nucleus rotation, see Supplementary Material (SM) Fig. S1.

We now explain how to correct telomere trajectories from cell motion, this is particularly important as we wish to explore the correlations  between  pairs of telomeres. To see how the drift of each different cell behaves with time ($t_i$), we computed the drift vector $\vec{R}(t_i)=(R_{x}(t_i),R_{y}(t_i))$, where each component is the respective  average center of mass: $R_{x}(t_i)= \sum_{j=1}^{M}x^{j}(t_{i}) /M$, and $R_{y}(t_i)= \sum_{j=1}^{M}y^{j}(t_{i}) /M$, (where $j=1,\ldots, M$ and $M$ is the number of tracked telomeres in the nucleus).  The relative magnitude of the drift vector $\vert \vec{R}(t)\vert - \vert \vec{R}(t_{0})\vert$ as a function of time for WT and KO cells is shown in Fig.~S4~\textit{a-b} in SM, here $\vert \vec{R}(t_i)\vert =\sqrt{R_{x}^{2}+R_{y}^{2}}$. The relative drift vector indicates the translational drift of the nucleus. In all cases the relative magnitude of the drift vector follows a smooth nearly linear behavior with time. 
More precisely, for thirty one different cells analyzed (19 WT and 12 KO), we observe that 12 of them (38\%) have almost zero drift and the others cells exhibit a quasi-monotonous trend with average slope  $\sim \pm 0.045 \mu m /s$.

To inspect the behavior of the rotation of the cell, for all the trajectories inside a given nucleus, we computed at each time frame the average rotation: $\phi(t_{i})= \sum_{j=1}^{M} \theta^{j}(t_{i},\Delta t) / M$, with $\theta^{j}(t_{i},\Delta t)$ the relative angle between successive vectors defined by Eq.~\eqref{eq:theta}. Here $\phi(t_{i})$ indicates at each time frame the average rotation around the center of mass of all telomeres in the cell. Fig.~S4~\textit{c-d} shows the rotation angle along the trajectory for few cells in both WT and KO cells. As one can see, the rotation is very small in most of the cases. 

To conclude these rather smooth/stable drift and small rotation angles, were corrected by the mentioned drift subtraction and rotation correction procedures, as one can find for other similar experimental set-ups~\cite{zhang2022}. A complete detail of our techniques are described in the Appendix  and in the SM.
\section*{Results}
\subsection*{\label{sec:22}\textbf{ Tracking telomeres in WT and KO cells}}
We tracked simultaneously fluorescently tagged telomeres in WT and KO MEFs cell lines, for a total measurement time  of  either $T=  925$ $s$ or  $1850$ $s$, as mentioned we tracked telomeres  in 19 WT cells and 12 KO, see Appendix~\ref{sec:42}  for  details.  Typical telomere trajectories are shown in Fig.~\ref{fig:micrograph}~\textit{c-d)}  for WT   and  KO cells. As mentioned above, while in WT cells  telomeres are mainly confined in  regions of typical length $\sim 100-200$ nm, in KO cells telomeres cover significantly larger distances up to $ \sim 3 \mu$m.  
Strikingly, we note that in KO cells, the diffusion not only spans a significantly larger volume, but one can notice that there are time-windows in which the motion looks highly directional (Fig.~\ref{fig:micrograph}~\textit{d}).
This variation in the directional dynamics, either caging with no preferential direction in WT cells or apparent unidirectional motion in KO cells, can be attributed to the respective differences in the nuclear structure of each type of cells. This difference can be characterized by the turning angle statistics, as explained below.

It was lately found that the dynamics of gene loci in WT cells is sub-diffusive, and it can turn into normal-like diffusion in KO cells ~\cite{Bronshtein2015}. It was also suggested that Lamin-A by itself, or together with the histone proteins ~\cite{melcer2012,prokocimer2009} forms intra- and inter-chromosomal cross-links of DNA,
thereby turning the free long chromosome polymer into a gel-like structure, which  naturally exhibits highly restricted dynamics in space and time. Moreover, it was shown that by transfecting Lamin-A back to these cells, the sub-diffusive nature of the chromatin dynamics is reestablished ~\cite{Bronshtein2015}. It demonstrates that Lamin-A is of major importance for maintaining a constrained diffusion. In addition, other nuclear structural proteins are known to take part in maintaining the chromatin structure, including the histon proteins that form the nucleosomes, cohesin and condensin that organize higher-order architecture such as the topologically associated domains, and factors such as CTCF, lap2$\alpha$ and BAF ~\cite{{Viva2019}}.
 Hereinafter, we explain how the presence/absence of Lamin-A protein renders a different behavior in the spreading of telomere trajectories, directed single particle dynamics, and correlation between pairs of particles.

\subsection*{\label{sec:23}\textbf{Laplace-like PDF of the telomeres displacement}}
After detrending telomere trajectories in the cell nucleus,  first we  analyze the dynamics of telomeres in WT and KO cells by calculating  the distribution of the traced displacements over different time intervals $\Delta t$. The distribution of displacements for  WT  telomeres was previously studied in ~\cite{Stadler2017}. Here we ask how does Lamin-A depletion modify this distribution and show that it has a dramatic effect on the scale of the motion.
However the distribution of displacements, as a stand alone does not yield insights on the coherent motion and correlations amongst telomeres, features that  will be treated later.

\begin{figure*}
\centering
\includegraphics[width=0.9\textwidth]{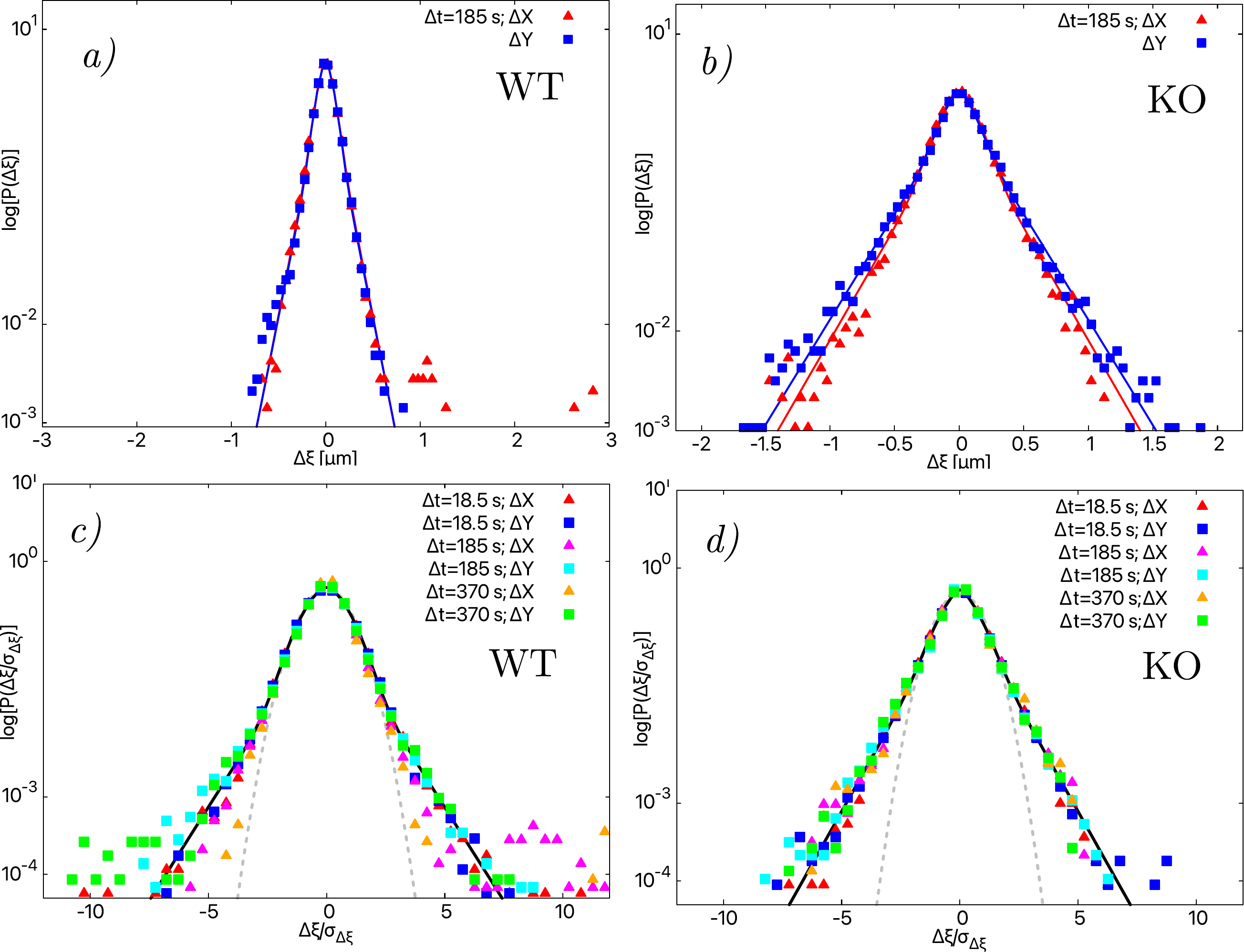}
\caption{Distribution of telomere displacements $\Delta \xi = \lbrace\Delta X, \Delta Y \rbrace$ (red triangles and blue squares in correspondence) WT \textit{a}) and KO \textit{b}) cells, with $\Delta t=185$ $s$. \textit{c-d}) Same data for normalized displacements $\Delta \xi / \sigma_{\Delta \xi}$, for WT and KO cells respectively for $\Delta t= \lbrace 18.5, 185, 370 \rbrace$ $s$, and X-Y directions (colored triangles x-coordinate and colored squares y-coordinate). Here $\sigma_{\Delta \xi}$ is the respective standard deviation. For comparison in each case Gaussian statistics is shown in gray dashed lines.  In all cases for any time scale, $\Delta \xi$ and cell type the positional PDF collapses into the same plot given by the model of a superposition of a Gaussian center and exponential tails Eq.~(1) (solid  lines). Nonetheless, the spreading of the positional PDF is  larger for telomeres in KO cells compared with WT cells. For WT cells, the total number of cells analyzed was:  $\cal{N}=19$, with 597 total number of tracks. For KO: $\cal{N}=12$ with 376  total number of trajectories. The exponential weights $w$ for WT and KO cases are shown in the Tables I and II in the SM.  }
  \label{fig:Propa}
\end{figure*}

The  displacements along the $x$ and $y$ axes are defined as  $\Delta X= X(t+\Delta t)- X(t)$ and $\Delta Y= Y(t+\Delta t)- Y(t)$; with $t=n\ast\Delta$ and $\Delta \xi = \lbrace \Delta X,\Delta Y \rbrace$.  Fig.~\ref{fig:Propa}  shows the  positional probability density function (PDF) in semi-log scale for telomeres in WT  and KO cells  with $\Delta t=10\Delta=185$ $s$, see  panels \textit{a} and \textit{b} respectively.   
For both cell types  we found that the positional PDF collapses to the same plot fitted by  a  model composed of a linear combination of Gaussian and  Laplace distributions (solid lines), which is defined as
\begin{eqnarray}
P(\Delta \xi)=(1-w)\frac{e^{- \frac{\Delta \xi^{2}}{4 \sigma^{2} }}}{\sqrt{4 \pi \sigma^{2} }} + w\frac{e^{-\frac{ \vert \Delta \xi \vert }{\lambda}. }}{2\lambda}, \label{eq:fitmod}
\end{eqnarray}
here $\lambda$ is the decay length of the exponential tails, $\sigma^{2}$ the variance of the Gaussian central part and $w$ the weight of the Laplace distribution. Distributions of displacements with exponential decaying tails have been thoroughly investigated both theoretically \cite{kob2007,hapca2009,chechk2017,bb2019,chuby2014,vittoria2018,LanG2018,WangCH_2020,eHSBB,hamdi2024} and experimentally ~\cite{leptos2009,wang2009,granick2012,spako2017,Janc2021,Sabri2020,aaberg2021,corci2023,hu2023}.
The analysis of residuals shown in Fig.~S8 in SM corroborates that Eq.~\eqref{eq:fitmod} is a good model  for describing telomere dynamics. We also tried to fit the data to a linear combination of two Gaussian curves. We found that the linear combination of a Gaussian and Laplace distributions gives significantly better results, see SM for details.

Interestingly,  previous studies ~\cite{Makhija2016} reported a Gaussian distribution for the displacements of telomeres in H2B-EGFP fibroblasts cells but  in the presence of  deformed nuclei and within a short time scale (3 min). Using micro-patterned substrates such nuclear deformation incurred in a reduction of Lamin-A/C levels. With such conditions telomere dynamics exhibit sub-diffusive behavior and Gaussian displacements, contrasting with the long time dynamics we report for telomeres in KO cells, \textit{i.e.} Eq.~\eqref{eq:fitmod} which presents a linear combination of Gaussian  and Laplace distribution for the displacements.        

There are several noteworthy observations about the distributions of telomere displacements (Fig.~\ref{fig:Propa}). First, we note that there are few displacements in Fig.~\ref{fig:Propa}a at large positive range. They correspond to only 0.13\% of the total set of points and can be considered as outliers. Secondly, note that there is a slight difference in the x and y coordinates in Fig.~\ref{fig:Propa}b. It results from only 3 out of 12 cells that are measured for KO cells (see Fig. S9 in SM). The nucleus of the cell is not always isotropic, so these results can occur naturally. Additionally, since cells and nuclei are non-identical objects, it is unsurprising that some of the cells are more anisotropic than others.

Fig.~\ref{fig:Propa}~\textit{c-d} show the distributions of the normalized displacements for $\Delta t= \lbrace \Delta, 10\Delta, 20\Delta\rbrace=\lbrace 18.5,185,370\rbrace$ $s$. The normalization is performed by dividing the displacements $\Delta \xi$ by the standard  deviation, defined as $\sigma_{\Delta \xi}=\sqrt{Var(\Delta \xi)}$. As one can see in both cell types and for all $\Delta t$, the PDF is well described by Eq.~\eqref{eq:fitmod} (black solid line).  For comparison in each case the Gaussian distribution  is shown as a gray dashed line,  further details of the fitting parameters are shown in the SM. The striking features are the exponential tails and the larger spread of the KO case, compared to the WT case (Fig.~\ref{fig:Propa}~\textit{a} -\textit{b}). 

Independently of the level of Lamin-A in the nucleus, the short displacements of the telomeres are described by Gaussian statistics and for large displacements the distribution has exponential tails, which coincides with the behavior of the Laplace distribution. A strong effect on the spreading of the displacements of telomeres was found when Lamin-A is completely depleted. 
From the fitting of Eq.~\eqref{eq:fitmod} to experimental data,  the  decay length obtained by the corresponding fitting, was $\lambda\sim 0.3$ $\pm 0.006$ $\mu m$ (Standard Error) for telomeres in KO cells when $\Delta t=370$ $s$, which is three times larger compared with  WT cells, $\lambda\sim 0.1$ $\pm 0.003$ $\mu m $ (Standard Error). The average total displacement $\langle \vert \Delta  \xi \vert \rangle =\langle\vert \xi(T) -  \xi(0) \vert \rangle$, behaves similarly with  $\langle \vert \Delta X \vert \rangle \sim 0.41 \pm 0.021$ $\mu m$ and $\langle \vert \Delta Y \vert \rangle \sim 0.45\pm 0.025$ $\mu m$  for telomeres in KO cells compared to $\langle \vert \Delta X \vert \rangle \sim 0.18 \pm 0.012$ $\mu m$ and $\langle \vert \Delta Y \vert \rangle \sim 0.17 \pm 0.006$ $\mu m$  for telomeres in WT cells.
Finally a  p-value calculation of the decay length in WT and KO cells,  with a null hypothesis of $\lambda_{WT}= \lambda_{KO}$  gives a very small p-value ($p \ll  0.001$).
Despite this difference in the spreading, for both cases we  found that the decay length $\lambda$ in Eq.~\eqref{eq:fitmod} grows as the square root of $\Delta t$, \textit{i.e.} $\lambda\propto \sqrt{\Delta t} $,
which is consistent with a linear MSD (see Fig.~S7 and Fig.~S11 in SM). This dependence of $\lambda$ was also found in other heterogeneous systems ~\cite{wang2009,Guan2014,miotto2021,Greco2022}. 

The information gained from Fig.~\ref{fig:Propa} is naturally an averaged quantity, namely one cannot pinpoint the origin of the exponential decaying tails, \textit{i.e.} Laplace diffusion ~\cite{StigLapl}.  Does it appear because different telomeres exhibit distinct types of motion?  Or is each trajectory heterogeneous, which means that there is an interplay of fast and slow motion? Fig.~\ref{fig:Propa} suggests  that  the Gaussian part of the PDF is associated with a slow dynamics phase  (for short distances), while the Laplace exponential decay is associated with fast motion (for long distances). A similar shape for the PDF of the displacements, namely a Gaussian center for short displacements and exponential-like for large ones,  was reported for probe particles (silica beads) in a cross-link actin network with myosin motors ~\cite{Toyota2011}. In ~\cite{Toyota2011} it is suggested that the Gaussian distribution in short-range displacements arises from thermal fluctuations and the exponential-like in the long-range displacements emerge from the active forces induced by the the molecular motors. The latter environment fits with the gel-like structure of inter-chromosomal cross links promoted by Lamin-A where telomeres move with different modes of motion, although the specific nature of this difference in the dynamics of telomeres  needs a further discussion. In the section Model, we show how these slow and fast phases are related to caging, pure diffusion and directed motion, and in the discussion  we go further with the biological explanation of such dynamics.

\subsection*{\label{sec:24}\textbf{Caged and Directed motion of telomeres inside the nucleus}}
\begin{figure}
\centering
\includegraphics[width=0.4\textwidth]{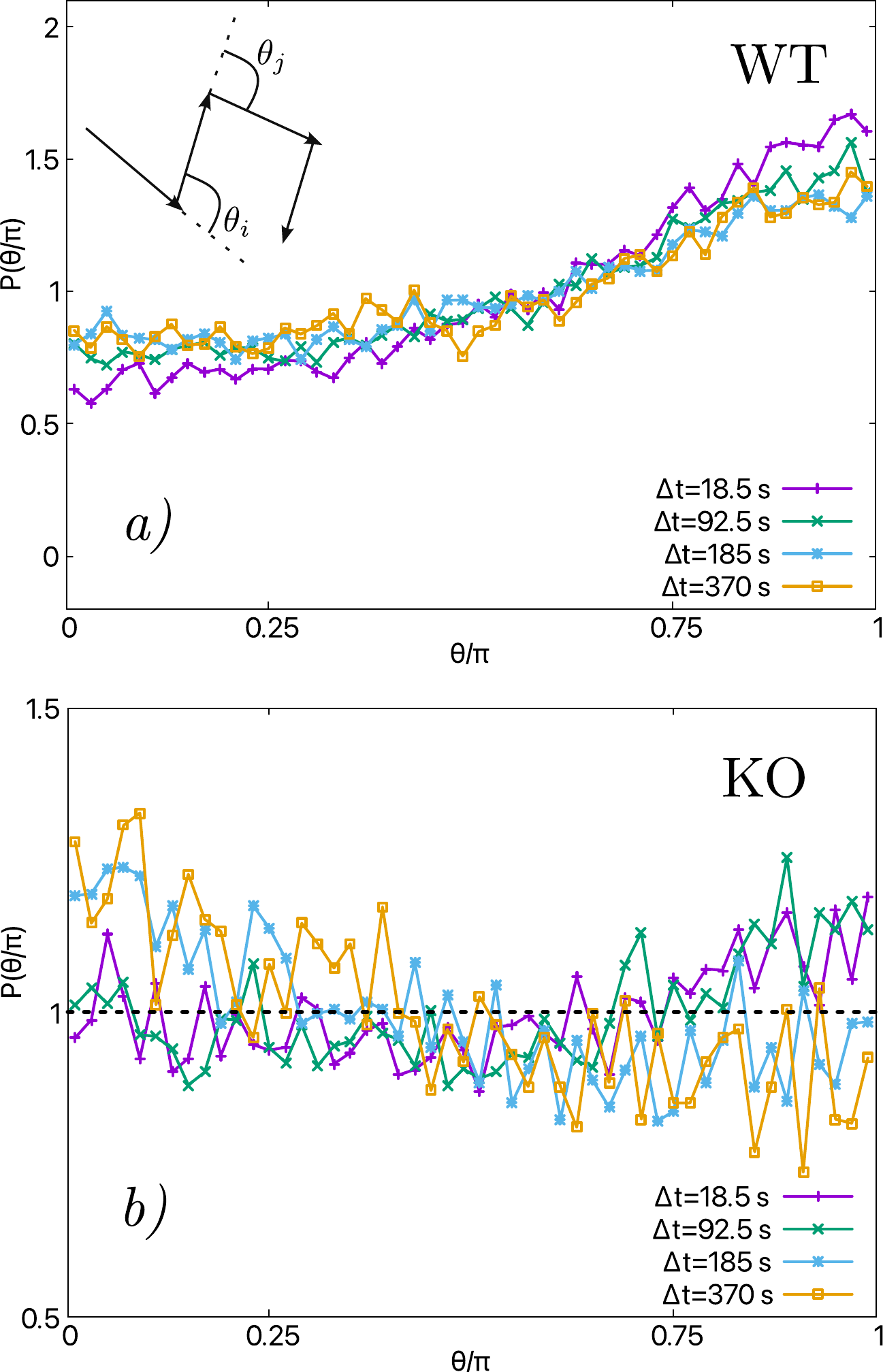}
   \caption{
    Probability distribution of the deflection angle along telomeres trajectories as a function of the normalized angle $\theta/\pi$ for few different lag-times $\Delta t = \lbrace 18.5,92.5,185,370 \rbrace$ $s$. \textit{a}) WT cells, the distribution mode is at $\theta=\pi$, indicating anti-persistent diffusion. Inset: examples of relative angles  $\theta_{i}, \theta_{j}$  defined by ~Eq.\eqref{eq:theta}.  \textit{b}) KO cells, the distribution mode is at $\theta=\pi$ for $\Delta t < 185$ $s$ and for larger time intervals $\Delta t > 92.5$ $s$ the peak of  $P(\theta / \pi)$ is at $\theta=0$,  indicating a transition from caging to   directed motion.  The black dashed line, represents the corresponding angular distribution for a random walk (a uniform distribution). For WT cells:  $\cal{N}=19$, with 597 total number of tracks. 
   And for KO: $\cal{N}=12$ with 376  total number of trajectories.
   }
  \label{fig:RelAng}
\end{figure}

The positional PDF shown above emphasizes the spatial and temporal scales of telomeres in the nuclei of WT and KO cells,  but  the  directional trajectories that only appear in short time-windows are averaged out. In order to expose its nature, we use a
 statistical tool that calculates  the relative angle between each two  points along the particle trajectory separated by $\Delta t$. Namely the absolute value of the turning angle which is denoted by $\theta(t;\Delta t)$ ~\cite{Burov2013,Schneider2015,Schneider2017,Huang2022,2024Krapf}, see inset in Fig.~\ref{fig:RelAng}~\textit{a}. The relative angle   for any  $\Delta t$   is defined as: 
\begin{eqnarray}\label{eq:theta}
\cos \theta(t,\Delta t)=\frac{\Delta\vec{X}(t;\Delta t) \cdot \Delta\vec{X}(t +\Delta t;\Delta t)}{\vert \vert \Delta\vec{X}(t;\Delta t) \vert\vert \,\,\ \vert \vert \Delta\vec{X}(t	+\Delta t;\Delta t) \vert\vert},\nonumber \\ 
\end{eqnarray}
with  the vector displacements $\Delta\vec{X}(t;\Delta t)=\Big(X(t+\Delta t)-X(t), Y(t+\Delta t)-Y(t)\Big)$  and $\vert \vert \cdot \vert \vert$ the Euclidean norm.

We compute for each $t$ the relative angle between successive displacements within different lapse-times $\Delta t= \lbrace 18.5, 92.5, 185, 370  \rbrace$ $s$  and calculate the respective distribution of the  relative angle. 
In persistent trajectories a particle tends to move in the same direction, exhibiting small values of $\theta(t;\Delta t)$. On the other hand, in anti-persistent tracks a particle flips its direction of  movement, remaining caged, and $\theta(t;\Delta t)$ has values close to $\pi$ ~\cite{Burov2013,MoernerSpako2012,Sadegh2017}.

Fig.~\ref{fig:RelAng}~\textit{a} shows the relative angle distribution  for telomeres  in WT cells, as a function of the normalized variable $\theta/\pi$.  We see that for any time scale, the relative angle distribution has a peak at $\pi$ and hence the displacements are negatively correlated, see the autocorrelation function of the displacements in  Fig.~S12~\textit{a}-\textit{b} in SM. Therefore the process is anti-persistent, implying caged dynamics ~\cite{MoernerSpako2012,Burov2013,Sadegh2017}.

In contrast, for KO cells $P(\theta)$ for short times $\Delta t <185$ $s$, it  has a maximum at $\theta=\pi$ and for larger time scales, $\Delta t > 92.5$ $s$, it has a peak at $\theta=0$, see Fig.~\ref{fig:RelAng}~\textit{b}. This implies that there is a transition from caging within short time scales to  directed motion for long time intervals. This persistent behavior in telomeres moving in  KO cells is translated into  trajectories like those displayed in Fig.~\ref{fig:micrograph}~\textit{d}, where telomeres tend to continue in the same direction for a significant time window.   

There is a striking difference between the behavior of $P(\theta)$ for both WT and KO cells (Fig.~\ref{fig:RelAng}) and the case of regular Brownian motion, \textit{e.g.} a random walk.
For Brownian dynamics  the distribution of the relative angle is uniform in the whole range of $\theta$, irrespective of $\Delta t$~\cite{Burov2013,Sadegh2017}.
Indeed, every movement is uncorrelated with the previous one, so every direction is equally probable (for any time scale).
The non-uniform profiles of $P(\theta)$ for the WT and KO  configurations reveal the temporal structure of the motion.
For the KO case within time scales of $100$ $s$ $ \lesssim t \lesssim 370$ $s$, the motion is directed as the particle has a preference to travel in the previously explored direction and for time scales of  $ t \lesssim 100$ $s$ telomeres exhibit anti-persistent behavior. On the other hand, WT dynamics shows the presence of caging  at any time scale (a peak for the relative angle distribution at $\theta=\pi$, see Fig.~\ref{fig:RelAng}~\textit{a}).

\begin{figure}
\centering
\includegraphics[width=0.4\textwidth]{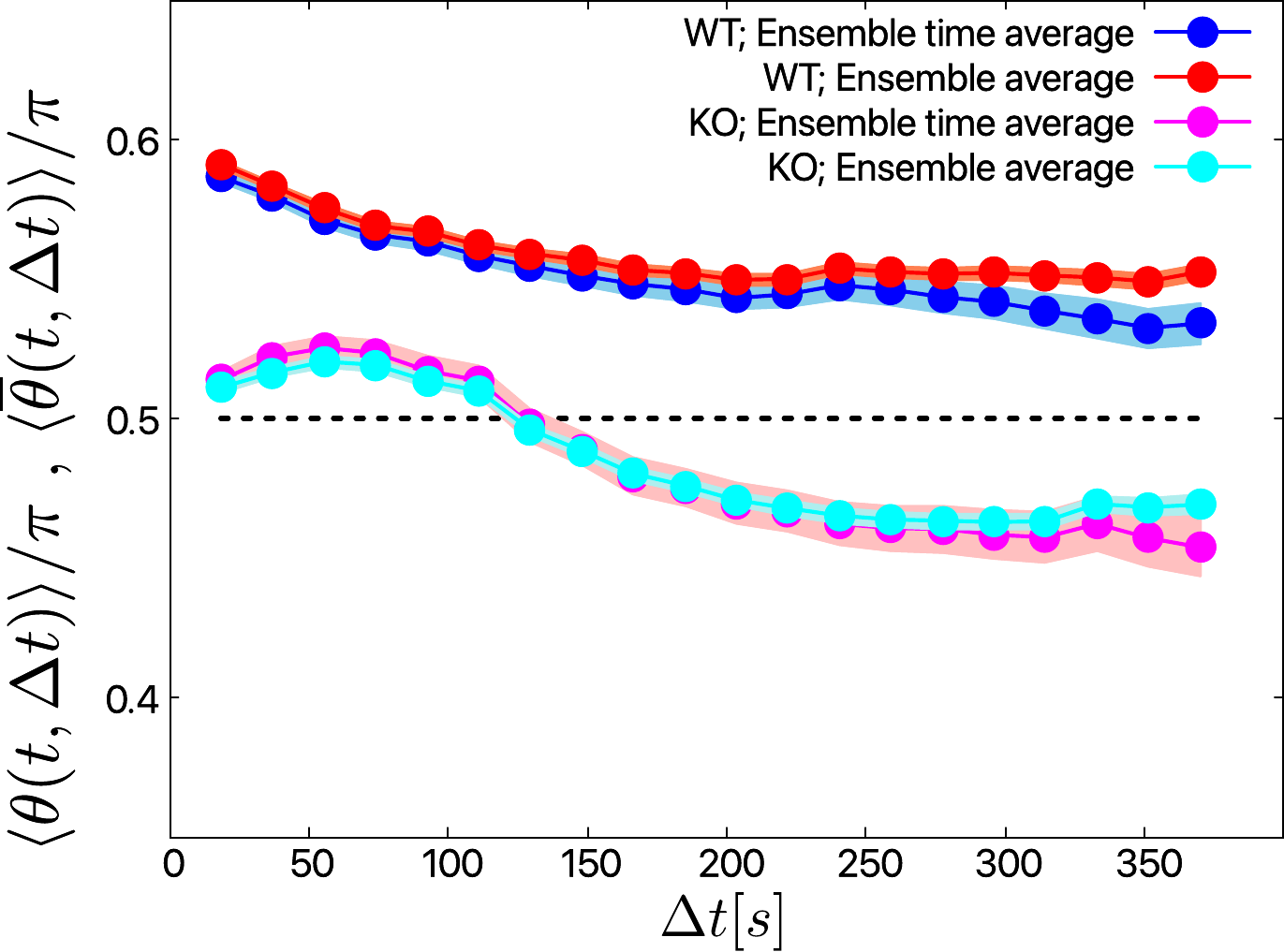}
\caption{
  \textit{Ensemble} average relative angle  ($\langle \theta(t,\Delta t)\rangle /\pi$) and \textit{ensemble} time average relative angle per trajectory $\langle \overline{\theta}(t,\Delta t)\rangle /\pi$ as a function of $\Delta t$. For WT cells the \textit{ensemble} average (red circles) and the \textit{ensemble} time average (blue circles) for all $\Delta t$ are above the corresponding values of a random walk, \textit{i.e.} 0.5 (black dashed line), indicating dominance of anti-persistent motion or caging.  For KO cells  the \textit{ensemble} average (cyan circles) and the \textit{ensemble} time average per trajectories (magenta circles) are initially for $\Delta t <130$ $s$ above 0.5 and then for time intervals $370$ $s$ $>\Delta t>100$ $s$ are below 0.5  which implies presence of persistent behavior. We employed for WT cells:  $\mathcal{N}=19$, with 597 total number of tracks. 
  And for KO: $\mathcal{N}=12$ with 376  total number of trajectories.  Shadowed areas denote the standard error of the ($SEM=SD / \sqrt{n}$).
}
  \label{fig:FTHC}
\end{figure}

The difference between caging and directed motion, becomes even clearer  when we observe the evolution of the  average relative angle  as a function of the sampling time (Fig.~\ref{fig:FTHC}).  For trajectories of telomeres in WT cells, the \textit{ensemble} average relative angle ($\langle \theta(t,\Delta t) \rangle /\pi$)    and the \textit{ensemble} time average of the relative angle ($\langle \overline{\theta}(\Delta t) \rangle /\pi$) are displayed in red circles and blue circles respectively. For tracks in KO cells, the respective averages are shown in cyan and magenta circles, the black dashed line indicates the average relative angle corresponding to a random walk, \textit{i.e.} $\langle \theta(t,\Delta t) \rangle /\pi=0.5$. Following results shown in Fig.~\ref{fig:RelAng}, in WT cells we observe caging dynamics for all measurement times. We observe a reduction of the average relative angle of 10\% from the short-time gap value to the long-time gap. For KO cells initially for  $ \Delta t <130$ $s$, the average relative angles are above 0.5, indicating caging. Then for $\Delta t > 130$ $s$,  the average relative angles are always less compared with the random walk reference value achieving  a reduction of 10\%, implying an increase in persistent motion along trajectories.
Finally, from Fig.~\ref{fig:FTHC} we observe that ergodicity is not satisfied in the time and \textit{ensemble} average relative angle (as it is also broken in the time and \textit{ensemble} MSD shown in Fig.~S7).  This ergodicity breaking is natural since the cell nucleus is a crowded and heterogeneous environment which induces this difference ~\cite{barkaiGarini2012}.    
 For different cell types (WT and KO), time and \textit{ensemble}  averages, a chi-square test against the constant value 0.5, accounting for the standard error of each data point (SEM), yields in all cases  small p-values \textit{i.e.}  $p \ll 0.001$, indicating a highly significant deviation from uncorrelated motion.

We conclude that the relative angle statistics  provides strong evidence for the existence of mechanisms that trigger both  caged and directed motion inside the cell nucleus, and this difference in the dynamics depends on the presence/absence of Lamin-A protein. Next we study how these directional dynamics are correlated on pair of telomeres in WT and KO cells.

Finally, other metrics like the asymmetry coefficient has been also employed to determine the level of directionality of chromatin sites ~\cite{iida2022}. While the asymmetry coefficient collapses directionality into a single forward–backward ratio based on arbitrary angle bins, the turning-angle method uses the entire angular distribution between successive steps. This makes it more sensitive to subtle differences in persistence or caging and provides a clearer, physically interpretable description of telomere motion.

\subsection*{\label{sec:25} \textbf{Pair correlation for telomere dynamics}}
 We now focus on the correlated motion of pairs of  telomeres, and show that the motion cannot be considered as mutually independent. In other words we search for the spatial correlation length; a scale on which dynamics is correlated.
We compute the pair correlation of telomeres as follows: for a pair of telomere tracks labeled as telomere $A$ and telomere $B$, in the same nucleus, we measure the initial separating  distance  $d_{AB}=\sqrt{(x^{A}(t_{1})-x^{B}(t_{1}))^{2}+(y^{A}(t_{1})-y^{B}(t_{1}))^{2}}$. We then compute the correlation function: 
\begin{eqnarray}\label{eq:Corr}
\beta_{A,B}(\Delta t, d_{AB})=\frac{1}{N}\displaystyle \sum \limits _{n=1}^{N}  \hat{\Delta X}_{A}(t_{n},\Delta t)  \cdot  \hat{\Delta X}_{B}(t_{n},\Delta t)\nonumber,\\
\end{eqnarray}
with $N$ the number of sampling points in each trajectory, $t_{n}=n*\Delta$, $\hat{\Delta X}_{A}(t_{n},\Delta t)=\overrightarrow{\Delta X}_{A}(t_{n},\Delta t)/\vert \vert\overrightarrow{\Delta X}_{A}(t_{n},\Delta t) \vert \vert$, where $\overrightarrow{\Delta X}_{A}(t_{n},\Delta t)= \overrightarrow{X}_{A}(t_{n}+\Delta t)- \overrightarrow{X}_{A}(t_{n})$ and $\overrightarrow{X}_{A}(t_{n})=(x^{A}(t_{n}),y^{A}(t_{n}))$.  We compute the \textit{ensemble} average $\langle \cdot \rangle$  of Eq.~\eqref{eq:Corr}  per cell type and bins for different values of $d_{AB}$ (see details in SM). Then, for a fixed value of $d_{AB}$, the average per cell type (WT or KO) is performed. For uncorrelated random walkers the pair correlation has zero values for any  $d_{AB}$ and $\Delta t$, as shown in Fig.~S14 in  SM.
 
\begin{figure}
\centering
\includegraphics[width=0.5\textwidth]{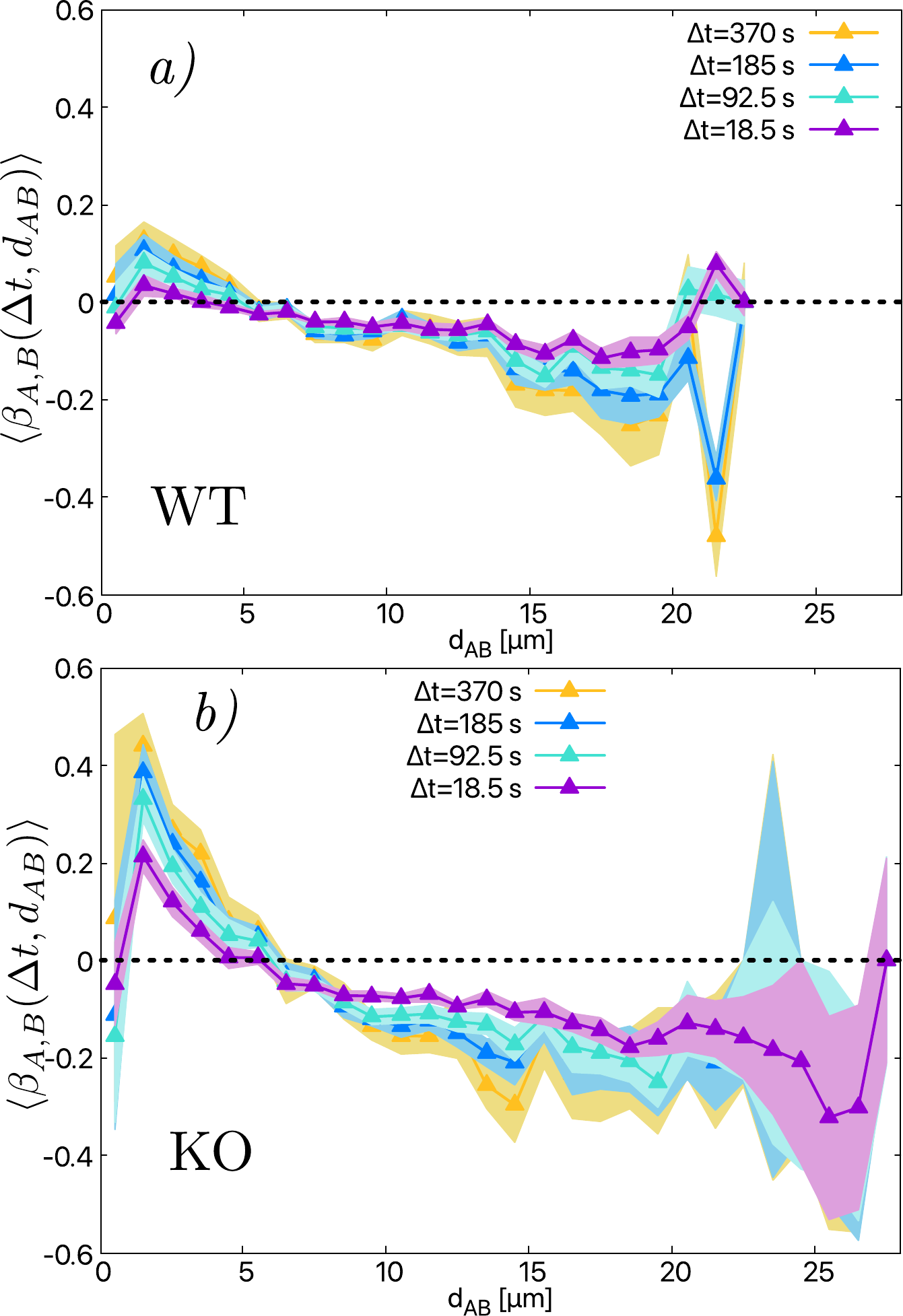}
   \caption{  Pair correlation of telomeres as a function of their initial separating distance. \textit{a}): For telomeres diffusing in WT cells, corresponding \textit{ensemble} average of the pair correlation of telomeres defined by Eq.~\eqref{eq:Corr}, for different measurement times $\Delta t = \lbrace 18.5, 92.5, 185, 370\rbrace$ $s$  (color triangles).  \textit{b}): The same as above but for telomeres in KO cells. For WT cells there are  small positive correlations of $\sim 0.07 $ at a separation distance of $\sim 1.5\mu m$ (for all time-scales) and a small negative correlations for large distances. For KO cells the pair correlation exhibits large positive values (maximum of 0.44) at short telomere distances of $\sim 1.5 \mu m$, and also negative correlation  at larger distances.
   This indicates that for both cell types the same correlation pattern appears. Nevertheless, it is significantly more pronounced in KO cells relative to WT cells. We employed for WT cells:  $\cal{N}=19$, with 597 total number of tracks. 
   And for KO: $\cal{N}=12$ with 376  total number of trajectories. Shadowed areas denote the respective $SEM$.
    }
  \label{fig:BCorr}
\end{figure}

Fig.~\ref{fig:BCorr} presents the pair-correlation of telomeres for different time-scales (Eq.~\eqref{eq:Corr}) as a function of the initial spreading distance between them. As one can see, in Fig.~\ref{fig:BCorr} for both cell types when telomeres are initially close to one another, the pair correlation is positive which means that they move coherently in the same direction. This effect is clearly more pronounced for KO cells, where the pair correlation exhibited significant positive values, for which the maximum attained is at 0.44. Particularly telomeres in WT cells exhibit regions of $\sim1\mu$m in extent where the pair correlations are positive, see Fig.~\ref{fig:BCorr}~\textit{a}. For KO cells the extent of positive correlations is $\sim 4\mu$m, see Fig.~\ref{fig:BCorr}~\textit{b}. Interestingly, in both cell types when telomeres are far away from each other inside the same cell, they move coherently in the opposite direction. This negative correlation can be the result of local external forces acting on the nucleus and locally modifying its shape, or nucleus breathing ~\cite{kepten2015uniform}. Remarkably we found that irrespectively the presence of Lamin-A, these pair correlations exhibit the same shape in both cell types, but the correlation level is different. Given the above observations, we developed a physical model for telomeres dynamics that can explain these results, paying specific attention to the caging and directed motion dynamics.
To evaluate whether the average pairwise correlations differ significantly from zero for each lag time $\Delta t$, we performed a chi-square test against the constant zero line, explicitly accounting for the standard error of each data point (SEM). For all lag times analyzed ($\Delta t = 18.5, 92.5, 185, 370$ s), the correlations were found to be significantly different from zero ($p \ll 0.001$) for WT and KO cells, confirming the presence of spatially correlated telomere motion.

Makhija et al.~\cite{Makhija2016} studied pairwise telomere correlations in Lamin-A/C–deficient cells under conditions of externally imposed nuclear deformation and reported a reduction in correlations. In our experiments, performed under native nuclear conditions without applied mechanical stress, Lamin-A depletion is instead associated with higher correlations compared to WT cells.

A reasonable, and perhaps the only explanation to the apparent difference in our results can be explained as follows, where we mainly refer to the highly constrained LP (long polarized) cells in Makhija's work. Chromatin, together with Lamin-A that forms cross links, is a relatively viscous and stiff, so that the nucleus primarily tends to move as a whole in response to cytoskeletal forces. Consequently, all telomeres share a similar displacement vector that leads to a very high pair-correlation value (0.65). This interpretation is further supported by the small nuclear area fluctuations measured in LP cells. Upon treatment with cytochalasin-D, which depolymerizes F-actin, cytoskeletal forces are reduced, leading to smaller nuclear fluctuations and therefore also reduced correlation. Importantly, this effect depends strongly on the constrained nature of LP cells, whereas CI (constrained isotropic) cells exhibit much weaker correlation coherence and even opposite trend when the cytoskeleton is modified. 
In Lamin-A deficient LP cells, the correlations are significantly reduced due to the loss of chromatin stiffness. This results in a softer chromatin structure so that external forces can act locally on the nucleus surface without moving it as a rigid body. It allows telomeres to move more freely thereby decreasing the pairwise correlations, even though the external cytoskeletal forces are still present. In our experiments the cells are not constrained, and the external forces acting on the nucleus are therefore much weaker. Under these conditions, the key factors distinguishing wild type from Lamin-A knockout cells are the chromatin structure and internal nuclear forces. In wild-type cells, the relatively rigid chromatin network suppresses large displacements induced by internal local forces, resulting in mostly small, weakly correlated motions. In contrast, in Lamin-A knockout cells, local nuclear forces can induce larger chromatin displacements, where nearby telomeres move along the force vector, which leads to increased correlations at short distances. Distant telomeres still remain weakly correlated without any major change.

\subsection*{\label{sec:26} \textbf{Model}}
Based on the experimental data of telomere dynamics in WT and KO cells, we developed a physical model that fits the main features exhibited by experimental data, like the exponential decay in the distribution of displacements and the relative angle distributions. All the parameters employed in our model are obtained from experimental data except for a few  free parameters dealing with the width of the fluctuations of the angular increments.

As reported in many single particle tracking experiments we see a wide fluctuation in behaviors of individual molecules ~\cite{Beni2011,Wero2017}. For instance  for telomeres in WT cells we found intermittency between caged and diffusive intervals, see Fig.~S20~\textit{a-f}) in the SM characterizing time series. This was deduced by computing the local convex hull (LCH) at each time frame and given a specific time window, see Fig.~S21 in the SM. Specifically at each time frame for the respective LCH we obtain the average largest diameter labeled as $\langle \mathcal{D}_{max} \rangle$. Then we define a criterion  for discriminating between the ``large jumps'' or diffusive phase and ``small jumps'' or caged  phase as following: If  $\mathcal{D}_{max}(t,\Delta t)>\langle \mathcal{D}_{max} \rangle$ the LCH contains ``large jumps'' (and its labeled as $L(t,\Delta t)=1$ ). In other case, $\mathcal{D}_{max}(t,\Delta t)<\langle \mathcal{D}_{max} \rangle$, the LCH contains ``small jumps'' (calling $L(t,\Delta t)=0$). Thus  we generate a dichotomous process $L(t,\Delta t)=\lbrace 0,1 \rbrace$ see cyan lines in Fig.~S20~\textit{d-f}). For telomeres in KO cells using a critical value for the relative angle at each time frame and within a specific time interval, we developed a similar decomposition of trajectories between persistent (${\cal P}(t,\Delta t)=1$) and non-persistent intervals (${\cal P}(t,\Delta t)=0$), see magenta lines in Fig.~S20~\textit{j-l}) and details in the Appendix characterizing the time series.

Therefore, for the dynamics of telomeres in WT cells, we suggest a modified Pearson random walk model.
Recall that for a Pearson random walk \cite{PEARSON1905} at  each time step, a random angle is chosen on the 2D plane and then the walker advances a distance $r$ on a straight line.
Accordingly we assume a two-state model of caging and pure diffusion, where intermittent switching occurs. During caging, the Pearson random walk is assumed in a circular cage of radius $r_c$ with reflecting boundary conditions. For the pure diffusive phase $r_c \rightarrow \infty$. We also employ different displacement lengths during different phases, namely $r_{S}<r_{L}$, with $r_{S}$ the displacement during caging and $r_{L}$ during pure diffusion. The duration of each of the two phases is randomly selected from an exponential distributions that are determined from the experimental data, see Fig.~S15 of the SM. The complete algorithm is presented in Materials and Methods~\ref{sec:48}. 

\begin{figure*}
\centering
\includegraphics[width=0.9\textwidth]{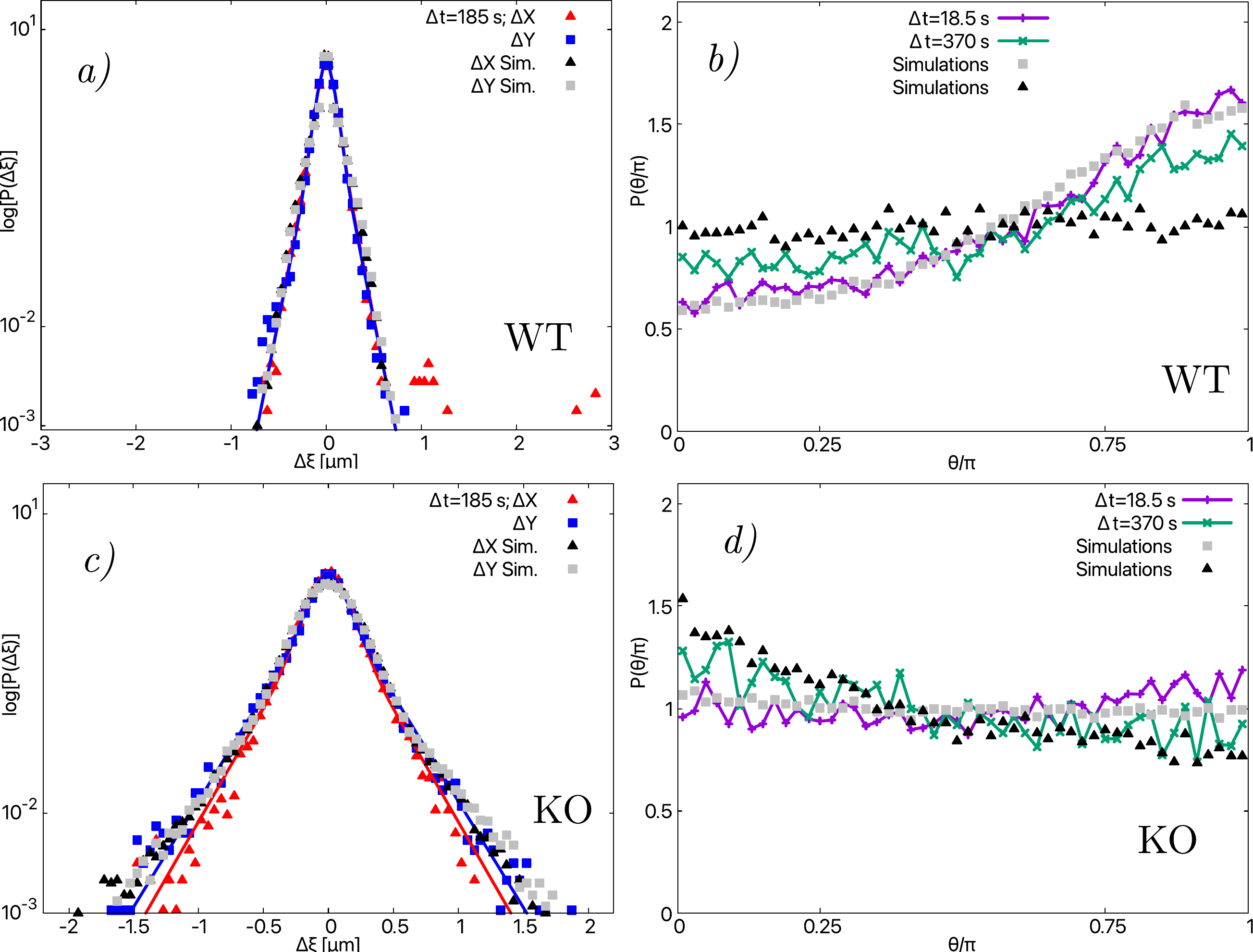}

\caption{Comparison between experimental data of telomeres in   WT and KO cells with simulations. \textit{a}) For telomeres in WT cells and simulation data of a two state model with caging and diffusive motion, distribution of displacements $\Delta \xi = \lbrace \Delta X, \Delta Y \rbrace$ with $\Delta t=185$ $s$. Experimental data is shown in red triangles and blue squares and simulation data in black triangles and gray squares respectively. \textit{b}) For telomeres in WT cells and simulation data of a two state model with caging and pure diffusive motion, relative angle distribution  experimental data in  color lines and simulation data in gray squares ($\Delta t=18.5$ $s$) and black triangles ($\Delta t=185$ $s$).  \textit{c}) and  \textit{d}) the same as in \textit{a}) and \textit{b}) but for telomeres in KO cells and simulation data from a two state model with Pearson and varying persistence random walks.  The parameters used in the simulations are specified in Materials and Methods~\ref{sec:48}.}
  \label{fig:FSim}
\end{figure*}

Fig.~\ref{fig:FSim} demonstrates a comparison of the experimental (Figs.~\ref{fig:Propa}, \ref{fig:RelAng}) and our simulation results (black triangles and gray squares). Fig.~\ref{fig:FSim} \textit{a},\textit{b} shows the comparison the displacements and relative angle in WT cells.
The agreement between theory and experiment is excellent for all the observed data.
The simulation data precisely follows the Gaussian distribution at short displacement and Laplace tails for larger ones Eq.~\eqref{eq:fitmod}. It also nicely follows the angular distribution data for all lag-times. Remaining statistics are displayed in Fig.~S17 in SM.

As explained before, previous works demonstrated the lack of chromatin cross links, which leads to directed dynamics for $\Delta t \leq 400 \Delta$ $s$ ~\cite{Bronshtein2015}. And as we show in Fig.~S20, in KO cells we found that telomere trajectories display a switching  between non-persistent and persistent intervals. Given this our model assumes intermittency between pure diffusion and directed dynamics. A similar dynamics was reported in chromatin sites of Chinese Hamster Ovary (CHO) cells ~\cite{levi2005}, where chromatin undergoes an apparently confined Brownian  motion alternating with moments of fast
curvilinear motion, these jumps were found to be  ATP-dependent. Thus the intermittently character of chromatin loci motion between passive and directed modes  is not restricted to the presence of Lamin proteins in the cell nucleus. Therefore  in the KO case the theory presents a two state model with a regular Pearson random walk and varying persistent motion. Our model works as follows. Initially, for each track we draw a random value for $\theta_{int}$ from a uniform distribution $U(0, 2\pi)$. We then build the trajectory with the first step equal to $(x_{1},y_{1})=(r\cos\theta_{int},r\sin\theta_{int})$ and each following step equal to $(x_{j},y_{j})=(x_{j-1}+r\cos\tilde{\theta}_{j},y_{j-1}+r\sin\tilde{\theta}_{j})$. Here $\tilde{\theta}_{j}=\tilde{\theta}_{j-1}+\tilde{\Delta \theta}_{j}$, where $\Delta \theta_{j}$  defines the level of persistence, and it  is drawn at each step from a uniform distribution $\tilde{\Delta \theta}_{j}\sim U(-\theta_{c},\theta_{c})$. Finally, a new $\theta_{int}$ is drawn again after $20\Delta=370$ $s$, see further details in the Appendix Simulation Models~\ref{sec:48}. By introducing $\theta_{int}$ at the beginning of each persistent phase,  we correlate the direction of successive segments of the  trajectory. This is done during time scales less than $370$ $s$, after this time a new $\theta_{int}$ is drawn and it is correlated in the successive segments during the same lapse of time.

We have the representative segments of trajectories, for instance: by correlating directed stages as above, $\theta_{c} \rightarrow 0$ implies alternating between straight lines and random walk paths, rendering persistent segments during a time scale  of $370$ $s$. And on the other hand,  when $\theta_{c}\geq \pi/2$ we have random walk behavior along the corresponding segments of the trajectories. The distribution of displacements and  one of the relative angles obtained from simulated trajectories are shown in gray squares and black triangles in Fig.~\ref{fig:FSim} ~\textit{c} and ~\textit{d} respectively. We find a good agreement with those obtained from experimental data. A Gaussian bulk with exponential decaying tails for $P(\Delta \xi)$ and a zero peaked angular distribution are recovered for $\Delta t > 100$ $s$, see further statistics in Fig.~S19 in SM. 

\section*{Discussion}

By examining the details of the dynamics along single telomere trajectories,  the following insights were exposed.
First, we identified a two-state behavior in both cell types WT and KO. In WT cells, the trajectories switch between caged and Brownian normal diffusion, while in KO cells they switch between pure diffusion and directed motion. 
Second, we found that in both cell types the dynamics lead to Gaussian statistics for small displacements and Laplace (exponential tails) statistics at large displacements. Our analysis revealed that the Gaussian statistics result from caging in WT cells and free diffusion in KO cells. Additionally, the Laplace tails results from the large displacements of the free diffusion in the WT cells and by the directed motion in KO cells. Thirdly, we note that when the dynamics of WT and KO cells is compared, it is clear that the directional motion observed in KO cells is suppressed by the effect of Lamin-A in WT cells.
It also explains the difference in the large spatial spread of the motion in KO cells versus small spread in WT cells.

Interestingly for trajectories in KO cells, we found persistent dynamics for times scales $100$ $s$ $ <\Delta t < 370$ $s$. For this case, if after each switching between the phases of motion the telomere ``forgets'' the directionality of the previously visited directed stage, the observed directionality is expected to disappear, in the sense that a uniform distribution of the relative angle describes such dynamics. Instead, the observed behavior of the relative angle PDF shows that the direction is persistent.

Fig.~\ref{fig:BCorr} shows the same interesting structure of pairwise correlations between WT and KO trajectories. Notably, for KO cells the magnitude of the correlations is larger, as was for the size of the displacements and the level of persistence along trajectories. This enhancement further supports the view that Lamin-A depletion amplifies coordinated chromatin motion within the nucleus.

\textit{Active forces}. Because the cells we analyzed are in interphase, we propose that the high levels of transcriptional and loop-extrusion activity characteristic of this phase contribute to the directed forces revealed in our measurements. These active forces act in both WT and KO cells, but the presence of Lamin-A normally constrains chromatin and dampens their effect on telomeres motion. In its absence, as in our KO cells, these forces become more apparent and likely contribute to the enhanced persistence and coherent behavior observed in telomere trajectories. Similarly, cytoskeletal forces ~\cite{Makhija2016,Stadler2017} may also contribute to these dynamics, fully disentangling the relative contributions of transcription ~\cite{Zidovska2013}, loop extrusion~\cite{samejima2025}.  Cytoskeleton-driven activity lies beyond the scope of this study and remains an important direction for future work.

{\em Laplace like tails}. The exponential decay of the PDF in the displacements of telomeres that was observed in this study and in ~\cite{Stadler2017}, has also been  observed in a several number of systems from glassy systems ~\cite{kob2007}  to diffusion of molecules within living cells~\cite{leptos2009,wang2009,granick2012,spako2017,Janc2021,Sabri2020,aaberg2021,corci2023} and tracers in colloidal suspensions \cite{Weeks2000,Kegel2000,WWEEKS2002,Yael2020,Lav2021}.  Laplace diffusion has been theoretically  modeled, by  diffusing diffusivity models ~\cite{hapca2009,chuby2014,chechk2017,vittoria2018,LanG2018,WangCH_2020,eHSBB},  spatial quenched disorder ~\cite{Liang2018,Liang2019,Postnikov2020,Pozo2021}, and interacting particle approaches ~\cite{Flavio2019,HSB2020,Sfluvio2021,Nampoothiri_2022}. It was shown that Laplace exponential statistics for the decay of positional PDF attains similar universal features as the Gaussian behavior of the PDF for the central part of the distribution~ \cite{bb2019,eWBB}. In this study, we showed for KO cells (via numerical simulations) that a correlated run and tumble model also leads to Laplace exponential decay. Nevertheless, the usual run and tumble models ~\cite{ST2012,Zabu2014,Fedotov2022}, as they traditionally have been developed for explaining the dynamics of bacteria, lack the long time correlations along trajectories that we have employed.   

{\em Directed motion in other systems and models}. Directional dynamics appears not only for run and tumble models~\cite{ST2012,Zabu2014,Fedotov2022}, but also for active Brownian particles ~\cite{SchehrMajumdar2018,Majumdar2020,Kundu2020}  and L\'{e}vy walks~\cite{DenKlaft2015,Kanazawa2020,wangY2020,Han2021,Mukh2021}.
The majority of these models lose directionality and do not establish an intrinsic correlation along single trajectories, as we suggest here for telomeres in KO cells. 
Another  popular approach to model active matter utilizes Langevin equations with active noises ~\cite{Volp2016,Jean2016}. 
The absence of spatial correlations (for the noise) in these models make them insufficient as a framework for our experimental system, as we have observed pairwise correlations between the dynamics of  telomeres. 
Interestingly in ~\cite{Kafri2021} the emergence of long-ranged active dynamics has been linked with the presence of disordered landscapes, where the concentration of active matter is non-uniform, in some cases fractal. Our results for KO cells are in agreement with the latter idea since we saw that the inherent directionality of telomeres tracks is quenched, and temporally correlated for hundreds of seconds.

Our detrending analysis shows that the raw drift displacement in both WT and KO cells is on the order of $4 \mu m$, whereas the detrended displacements are much smaller, approximately $\lambda = 0.1 \mu m$  for WT cells and $0.3 \mu m$ for KO cells. Thus, the global drift is roughly one order of magnitude larger than the true local telomere motion. As noted earlier, lamin-A knockout also increases telomere displacement by about a factor of three. While our detrending procedures follow standard single-particle tracking approaches, additional caution is required when extending the analysis to correlations among multiple simultaneously tracked telomeres.

Unexpectedly, when we study the directional correlations
of pairs of telomeres inside the same cell, we observe
that there is a general structure for such correlations in
WT and KO cells, \textit{i.e.} positive correlations for short distances
between telomeres and negative values for large
distances (Fig.~\ref{fig:BCorr}). The latter is important because
independently of the presence of Lamin-A, it suggests the
existence of long-ranged forces that create this coherent
motion in the nucleus.
Taken together, these observations indicate that Lamin-A not only shapes nuclear architecture and diffusion properties, as previously shown, but also regulates inter-locus correlations, effectively suppressing them in WT cells.

\section*{Appendices}
\section*{\label{sec:42}\textbf{Analysis of Trajectories}}
We perform SPT analysis  of fluorescently labeled telomeres  in  WT cells, where 597 trajectories of telomeres were analyzed,  504 tracks with $\Delta t=50\Delta=925$ $s$ and 93 with $\Delta t=100\Delta=1850$ $s$. In the case of telomeres in KO cells, 376 trajectories  were analyzed, 307 tracks with $\Delta t=50\Delta=925$ $s$ and 69 with $\Delta t=100\Delta=1850$ $s$. Particularly for WT cells and measurement time $T=50 \Delta$, we employed 14 different cells and on each one we tracked the following numbers of trajectories: cell \#1 25 tracks, cell \#2 35 tracks, cell \#3 38 tracks, cell \#4 25 tracks, cell \#5 35 tracks, cell \#6 48 tracks, cell \#7 25 tracks,  cell  \#8 39 tracks, cell \#9 35 tracks, cell \#10 62 tracks, cell \#11 39 tracks, cell \#12 39 tracks, cell \#13 23 tracks, and cell \#14 36 tracks  ( with a total of 504 trajectories). For WT cells and $T=100 \Delta$, we employed 5 different cells for which we trace: cell \#1 21 tracks, cell \#2 24 tracks, cell \#3 20 tracks, cell \#4 13 tracks, and cell \#5 15 tracks (with a total of 93 trajectories). In the case of  KO cells and measurement time $T=50 \Delta$, we used  8 different cells each one following: cell \#1 19 tracks, cell \#2 17 tracks, cell \#3 23 tracks, cell \#4 78 tracks, cell \#5 41 tracks, cell \#6 37 tracks, cell \#7 54 tracks, and cell \#8 38 tracks    (with a total of 307 trajectories) . And finally for KO cells and $T=100 \Delta$, we employed 4 different cells with: cell \#1 18 tracks, cell \#2 26 tracks, cell \#3 10 tracks, and cell \#4 15 tracks  (with a total of  69 trajectories).

In all cases a lag time of $\Delta=18.5$ $s$ was employed, and after performing the respective SPT, a nucleus drift and rotation correction were applied. 
For the correction of the nucleus drift, per cell at each multiple of the lag time $t_{i}=i\ast \Delta$, the center of mass of each coordinate was computed: $\langle x(t_{i})\rangle= \sum_{j=1}^{M}x^{j}(t_{i}) /M$, $\langle y(t_{i})\rangle= \sum_{j=1}^{M}y^{j}(t_{i}) /M$, $\langle z(t_{i})\rangle= \sum_{j=1}^{M}z^{j}(t_{i}) /M$ (where M is the number of telomeres in the nucleus). Thereafter each center of mass was subtracted forward in time (for $t_{k}>t_{i}$ with $k>i$) to the corresponding coordinate: $x^{j \ast}(t_{k})=x^{j}(t_{k})- \langle x(t_{i}) \rangle$, $y^{j \ast}(t_{k})=y^{j}(t_{k})- \langle y(t_{i}) \rangle$ and $z^{j \ast}(t_{k})=z^{j}(t_{k})- \langle z(t_{i}) \rangle$. Next  the nucleus rotation correction was performed as follows: per cell and for all the tracks inside at each frame of time we follow: when $i=1$ (the first frame of time) we compute $r_{d}=\sqrt{(x^{j\ast}(t_{1}))^{2}+ (y^{j\ast}(t_{1}))^{2}}$ and $\theta^{j}(t_{1})=acos(x^{j \ast}(t_{1})/r_{d})$, if $y^{j \ast}(t_{1})<0$, we make $\theta^{j}(t_{1})=-\theta^{j}(t_{1})$. Then we compute the average as   $\langle \theta (t_{1})\rangle= \sum_{j=1}^{M} \theta^{j}(t_{1}) / M$. When $i>1$, we compute $r_{1}=\sqrt{(x^{j\ast}(t_{i}))^{2}+ (y^{j\ast}(t_{i}))^{2}}$ and $r_{2}=\sqrt{(x^{j\ast}(t_{i-1}))^{2}+ (y^{j\ast}(t_{i-1}))^{2}}$. With $\theta^{j}(t_{i})=acos((x^{j\ast}(t_{i})\ast x^{j\ast}(t_{i-1})  + y^{j\ast}(t_{i})\ast y^{j\ast}(t_{i-1}) ) / (r_{1}r_{2})  )$ and if $asin((-x^{j\ast}(t_{i})\ast y^{j\ast}(t_{i-1})  + y^{j\ast}(t_{i})\ast x^{j\ast}(t_{i-1}) ) / (r_{1}r_{2})  ) <0 $, we make $\theta^{j}(t_{i})=-\theta^{j}(t_{i})$. Then the average is computed just when $r_{2}>0.5$ (neglecting trajectories close to the origin), as   $\langle \theta (t_{i})\rangle= \sum_{j=1}^{M} \theta^{j}(t_{i}) / \tilde{M}$, with $\title{M}$ the number of track which satisfy this criterion. Then as above for $t_{k}>t_{i}$ with $k>i$, the inverse rotation matrix was applied as: $x^{j \prime }(t_{k})=x^{j\ast}(t_{k})\cos(\langle \theta^{j}(t_{i}) \rangle)+ y^{j\ast}(t_{k})\sin(\langle \theta^{j}(t_{i}) \rangle)$ and $y^{j\prime }(t_{k})=- x^{j\ast}(t_{k})\sin(\langle \theta^{j}(t_{i}) \rangle)+ y^{j\ast}(t_{k})\cos(\langle \theta^{j}(t_{i}) \rangle)$. In this way the final coordinates $(x^{j\prime}(t_{k}),y^{j\prime}(t_{k}))$ have no trend or bias in rotation. The full algorithm is shown in SM.
\section*{\label{sec:46} \textbf{Characterizing the time series}}
{\em Using the relative angle time series.} To differentiate between periods of persistent and non-persistent motion on a single track in KO cells, we employed  the relative angle defined by ~\eqref{eq:theta} at an intermediate span of time $\Delta t=5\Delta$ and $\Delta t=10\Delta$.  Thus  from the time series of the telomeres without Lamin-A $\lbrace x(t),y(t) \rbrace$ with $t=n\ast\Delta$, first we obtain the time series on the angle $\lbrace \theta(t,\Delta t) \rbrace$.  For each time step we compare $\theta(t,\Delta t) $ with  a critical value named $C$. In this case $C$ defines a threshold of the angles, letting the separation between the persistent and non-persistent motions,  we define  the critical relative angle as $C=30^{\circ}$ ~\cite{Mukh2021}.

So the criterion for distinguish between persistent and non-persistent motion is the following: If $ \theta(t,\Delta t)  < C$ (there is persistent motion) and we label it as $ {\cal P}(t,\Delta t) =1$. Else if $ \theta(t,\Delta t) \geq C$ (the dynamics in non-persistent) and it is labeled as $ {\cal P}(t,\Delta t)  =0$. In this way we generate a two state time series which we call ${\cal P}(t,\Delta t)=\lbrace 0,1 \rbrace$ (see magenta lines in Fig.~S20). 

{\em Using local convex hull analysis.} A typical statistical tool for detecting the existence of different phases of transport with various  jump sizes is the local convex hull (LCH) analysis ~\cite{LanGreb2017,Sabri2020}. LCH technique is described as follows. For a data set with points $\lbrace \vec{X}_{1}, \vec{X}_{2}, \ldots, \vec{X}_{m} \rbrace$ its  convex hull ($CH$) is the set of all convex combinations which can be made with the data points, 
\begin{eqnarray}\label{eq:LCH}
CH=\Bigg \lbrace \displaystyle \sum \limits _{k=1}^{m} \alpha_{k}\vec{X}_{k} \Bigg \vert \displaystyle \sum \limits _{k=1}^{m} \alpha_{k}=1\Bigg\rbrace.
\end{eqnarray}

In the 2$D$ case  $\vec{X}\in \mathbb{R}^{2}$, $CH$ is the minimal convex polygon that encloses all the data points $\lbrace \vec{X}_{1}, \vec{X}_{2}, \ldots, \vec{X}_{m} \rbrace$. For employing CH in the time series of the telomeres, we used time windows $\Delta t=5\Delta,$ and $ \Delta t= 10\Delta$, such that at each $t=n\ast \Delta$ we define the corresponding  LCH given by ~\eqref{eq:LCH} on the data points $\lbrace \vec{X}_{t}, \vec{X}_{t+\Delta}, \vec{X}_{t+2\Delta}, \ldots, \vec{X}_{t+\Delta t} \rbrace$. Then for each LCH we compute the maximum diameter $\mathcal{D}_{max}(t,\Delta t)$, largest distance between the points $\lbrace \vec{X}_{t}, \vec{X}_{t+\Delta}, \vec{X}_{t+2\Delta}, \ldots, \vec{X}_{t+\Delta t} \rbrace$, of the corresponding polygon, see Fig.~S21. In this figure we show two typical LCH polygons  (using Eq.~(4) in the main text and within $\Delta t=5\Delta$) in red color solid line. Also in the same figure we show the Area $A$ (displayed in red shaded color) and the maximum diameter (black dashed lines) of the respective LCH polygon.

\section*{\label{sec:466} \textbf{Estimation of the waiting times}}
Representative examples of the  two state decomposition of trajectories in terms of the level of persistence (${\cal P}(t,\Delta t)$) for KO cells  and the relative jump size ($L(t,\Delta t)$) for WT cells,  is shown in Fig.~S20. 

In both cells types using the time series   of ${\cal P}(t,\Delta t)$ (magenta lines) and $L(t,\Delta t)$ (cyan lines), we extracted the waiting times corresponding to the persistent phase and non-persistent phase, labeled as   $\tau_{{\cal P}}$ and $\tau_{N{\cal P}}$ respectively. And the waiting times  $\tau_{L}$ and $\tau_{S}$  for the long jump phase  and the small jump one. In all cases the corresponding distributions follow a discrete exponential distribution, well fitted by the geometric distribution with mass distribution $P(\tau)=(1-p)^{\tau -1}p$, with $0<p\leq 1$ and $\tau=1,2,3,\ldots$, (see  Fig.~S15 in SM). Via this classification of trajectories  we obtained the mean waiting times for trajectories in WT cells, $\langle  \tau_{L} \rangle =5.8 \Delta < \langle  \tau_{S}\rangle =9.5\Delta $ using $\Delta t=5 \Delta$, and  $\langle \tau_{L} \rangle=9.2\Delta < \langle \tau_{S} \rangle=14.9\Delta$ employing $\Delta t=10 \Delta$. For trajectories in KO cells, $\langle  \tau_{{\cal P}} \rangle =1.6\Delta < \langle  \tau_{N{\cal P}}\rangle =6.7\Delta $ using $\Delta t=5 \Delta$ and  $\langle \tau_{{\cal P}} \rangle=2\Delta < \langle \tau_{{N{\cal P}}} \rangle=7.4\Delta$. 

\section*{\label{sec:48} \textbf{Simulation models}}
{\em Two state model with caging and pure diffusion}. In this case we define two waiting times, one for the caged with small length jumps called $\tau_{S}$ and other for the pure diffusive phase with larger length jumps named $\tau_{L}$. Both waiting times are sampled from the corresponding empirical distribution $P(\tau_{S})$ and $P(\tau_{L})$, see Fig.~S15~\textit{b}. These distributions  were obtained from the two state decomposition in terms of the jump size length of the trajectories and represented by the time series of $L(t,\Delta t)$ (see Fig.~S20). And both were fitted with the geometric distribution with mean values $\langle  \tau_{L} \rangle =9 \Delta < \langle  \tau_{S}\rangle =14.9 \Delta $, see SM. 

We consider equilibrium initial conditions, namely with probability $\langle  \tau_{S}\rangle/ [\langle  \tau_{S}\rangle+\langle  \tau_{L}\rangle]$ the particle starts from the caging phase and with probability $\langle  \tau_{L}\rangle/ [\langle  \tau_{S}\rangle+\langle  \tau_{L}\rangle]$ it starts from the pure diffusive one. Then the trajectory alternates between the two phases until a final time $T \ast N$, with $N$ the sampling time.

Particularly each phase is defined as follows. For the caging phase during a random time $\tau_{S}\ast N$ for each sub-step  we implement a  Pearson random walk, with $\tilde{\theta}_{i}\sim U(0,2\pi)$ and  $(x_{i},y_{i})=(x_{i-1}+r_{S}*\cos\tilde{\theta}_{i},y_{i-1}+r_{S}*\sin\tilde{\theta}_{i})$, and reflecting boundary conditions within a circular cage with radius $r_{c}=0.01\mu$m. For the pure diffusive phase, $r_{c}\longrightarrow \infty$, during a random time $\tau_{L}\ast N$, we sample $\tilde{\theta}_{i}\sim U(0,2\pi)$ and  $(x_{i},y_{i})=(x_{i-1}+r_{L}*\cos\tilde{\theta}_{i},y_{i-1}+r_{L}*\sin\tilde{\theta}_{i})$. Notice that the displacement length depends on the phase where the particle is, they satisfy $r_{S}<r_{L}.$

We sample the positions along the corresponding trajectory  each $N$ sub-steps. The simulations were done  using parameters close to the experimental ones. A final time of  $T=50$,  sampling time of $N=20$,  displacement length in the caging phase $r_{S}=0.005\mu$m, $r_{L}=0.02\mu$m in the pure diffusive phase and $\langle  \tau_{S} \rangle =9 \Delta < \langle \tau_{L}\rangle =15 \Delta $. We employ an \textit{ensemble} of 3000 different particles.

{\em Two state model with Pearson and  varying persistence random walks}.  
Similarly as for the model defined above we define a non-persistent Pearson random walk phase with waiting times $\tau_{N{\cal P}}$ and a varying persistence phase with $\tau_{{\cal P}}$. The corresponding waiting times are sampled from the distributions obtained from the two state analysis of the level of persistence, see Fig.~S15~\textit{d} in SM. 

For each trajectory we draw an angle $\theta_{c} \sim N(\bar{\theta},\sigma^{2}_{\bar{\theta}})$, with  $\bar{\theta}(t, \Delta t)$ the average relative angle and $\sigma^{2}_{\bar{\theta}}$ its respective variance. Also at the beginning $\theta_{int}\sim U[0,2\pi]$ is drawn.  We consider equilibrium initial conditions, namely with probability $\langle  \tau_{N{\cal P}}\rangle/ [\langle  \tau_{N{\cal P}}\rangle+\langle  \tau_{{\cal P}}\rangle]$ the particle starts from the Pearson random walk phase and with probability $\langle  \tau_{{\cal P}}\rangle/ [\langle  \tau_{N{\cal P}}\rangle+\langle  \tau_{{\cal P}}\rangle]$ it starts from the varying persistent one. Then the trajectory alternates between the two phases until a final time $T \ast N$, with $N$ the sampling time.

Thus for the Pearson random walk phase during a random time $\tau_{N{\cal P}}\ast N$ for each sub-step  we do (without restrictions of motion)  $\tilde{\theta}_{i}\sim U(0,2\pi)$ and  $(x_{i},y_{i})=(x_{i-1}+r*\cos\tilde{\theta}_{i},y_{i-1}+r*\sin\tilde{\theta}_{i})$. For the varying persistent phase we proceed as follows.  During a random time $\tau_{{\cal P}}\ast N$, for the first sup-step we make $\tilde{\theta}_{i}=\theta_{int}$ and we update the positions as $(x_{i},y_{i})=(r*\cos\theta_{int},r*\sin\theta_{int})$. Then for sub-steps $j>i$, we draw $ \tilde{ \Delta\theta }_{j}\sim U(-\theta_{c},\theta_{c})$, we make $\tilde{\theta}_{j} = \tilde{\theta}_{j-1} + \tilde {\Delta \theta}_{j}$ and $(x_{j},y_{j})=(x_{j-1}+r*\cos\tilde{\theta}_{j},y_{j-1}+r*\sin\tilde{\theta}_{j})$. In order to avoid correlations in the varying persistence phases for time scales larger that $20\Delta$, at each $20\ast \Delta$, $\theta_{int}$ is drawn again.  

Sampling the positions along the corresponding trajectory  each $N$ sub-steps. The respective parameters employed were $T=50$,  $N=20$,  $r=0.015\mu$m, $\langle  \tau_{{\cal P}} \rangle =2 \Delta < \langle  \tau_{N{\cal P}}\rangle =7 \Delta$, $\bar{\theta}=1.48$, $\sigma^{2}_{\bar{\theta}}=0.86$ and an \textit{ensemble} of 3000 different particles.

\section*{Author Contributions}
EB, YG and SB designed the research; YG, YH and WN performed the experiments; MHS and SB performed the data analysis; MHS performed the numerical simulations; MHS, SB, EB, WN and YG wrote the paper.

\section*{Declaration of Interests}
The authors declare no competing interests.

\section*{Data Availability}
The data supporting the findings of this study, including the telomere trajectory datasets analyzed in this work, are publicly available at: \url{https://github.com/mariohidalgosoria/Telomeres_Dynamics/tree/main/Data}.

\section*{Acknowledgments}

The support of Israel Science Foundation Grants  1614/21 (EB and MHS) and 2796/20 (SB) is acknowledged  as well as the Israel Science Foundation Grants  1902/12, 1219/17 and 2624/22 (YG, YH and WN) and the S. Grosskopf grant for Generalized dynamic measurements in live cells at Bar Ilan University.

\bibliography{rsc}


\section*{Supplementary Material}

An online supplement to this article can be found by visiting BJ Online at \url{http://www.biophysj.org}.

\end{document}